# Fermi Surface reconstruction and anomalous low temperature resistivity in electron-doped $La_{2-x}Ce_xCuO_4$

Tarapada Sarkar[1,2], P. R. Mandal[1,2], J. S. Higgins[1,2], Yi Zhao[1,2], Heshan Yu[3], Kui Jin[3], Richard L. Greene[1,2]

[1]Center for Nanophysics & Advanced Materials, University of Maryland, College Park, Maryland 20742, USA.
[2]Department of Physics, University of Maryland, College Park, Maryland 20742, USA.
[3]Beijing National Laboratory for Condensed Matter Physics, Institute of Physics, Chinese Academy of Sciences, Beijing 100190, China

We report ab-plane Hall Effect and magnetoresistivity measurements on $La_{2-x}Ce_xCuO_4$ thin films as a function of doping for magnetic fields up to 14T and temperatures down to 1.8K. A dramatic change in the low temperature (1.8 K) normal state Hall coefficient is found near a doping Ce=0.14. This, along with a nonlinear Hall resistance as a function of magnetic field, suggests that the Fermi surface reconstructs at a critical doping of Ce= 0.14. A competing antiferromagnetic phase is the likely cause of this Fermi surface reconstruction. Low temperature linear-in-T resistivity is found at Ce=0.14, but anomalously, also at higher doping. We compare our data with similar behavior found in hole-doped cuprates at a doping where the pseudogap ends.

Correspondence: rickg@umd.edu

The mechanism responsible for the high-temperature superconductivity in the cuprates, and the nature of the normal state from which it evolves, is a major unsolved problem in condensed matter physics. Most of the research on cuprates has focused on hole-doped materials, which are more numerous. However, the few examples of electron-doped cuprates offer many advantages for a possible solution to the high-Tc superconductivity problem. The doping phase diagram is much simpler for n-type cuprates. The superconductivity evolves from an antiferromagnetic (AFM) state without the mysterious "pseudogap" state found in the hole-doped cuprates [1,2]. Moreover, the critical magnetic field needed to suppress the superconductivity is much lower for electron-doped cuprates so that the fundamentally important non-superconducting ground state can be probed by experiment. In recent work [3] on $La_{2-x}Ce_xCuO_4$ (LCCO) a surprising linear in temperature normal state resistivity was discovered at low temperature (30mK to 10K) over a range of Ce doping. The strength of the T-linear resistivity was proportional to the superconducting transition temperature (Tc), which suggested that antiferromagnetic spin fluctuations were responsible for both. Theory suggests that AFM should end at a quantum critical point (QCP) and only at the QCP might a T-linear resistivity be found [4-6]. It is also thought that quantum fluctuations associated with a QCP can lead to superconductivity. In LCCO long range AFM ends at a doping near Ce=0.09[7] where no T-linear resistivity is found. Short range magnetism persists to higher doping, but where it ends is unknown [8]. Superconductivity exists over the approximate doping range 0.08 to 0.17, with conventional metallic (Fermi liquid) behavior at higher doping. These prior results raise several important questions of relevance to the origin of high-temperature superconductivity (HTSC) in the cuprates: 1) can short range magnetic order produce QCP-like behavior, and 2) can short range order cause a Fermi surface reconstruction. In this paper we present new low temperature transport measurements on electron-doped LCCO that show the answer to these questions is yes. These surprising experimental conclusions will need new theoretical ideas to reconcile them with the extended range of T-linear resistivity found previously [3].

Our new results on LCCO are also of significance in comparison with recent studies of hole-doped cuprates at very high magnetic fields. In particular, normal state Hall effect measurements done at fields up to 90T on hole-doped YBCO [9] and LSCO [10] have received much attention because they suggested that a Fermi surface reconstruction (FSR) occurs at a critical doping (p*) under the superconductivity (SC) dome. This critical doping is also where the mysterious pseudogap (PG) ends. It was found that a large Fermi surface at p > p* transitions to a small Fermi surface at p < p* corresponding to a Hall number $n_H$ which goes from 1+p to p. This recent transport work agrees with prior SI-STM [12] and other experiments [13] which suggested a FSR at a doping near 0.19 in hole-doped cuprates. Although the exact cause of the pseudogap is unknown the p* end point has recently been suggested to be related to the end of spiral AFM [14] or a novel "topological" phase transition [15]. As shown in Fig. 5 of this work, we find a very similar change in Hall number at our suggested FSR in LCCO at a critical doping (p*) of Ce=0.14. But, in our case it is almost certainly short range magnetism that ends at p*. Since the physics that drives the FSR and the SC is likely to be the same on both sides of the cuprate phase diagram, our results appear to be of considerable significance for a deeper understanding of the HTSC in the cuprates.

LCCO is unique among n-doped cuprates because it can be prepared in thin film form over a wider range of doping, in particular beyond the superconducting dome. However, some prior work on other n-type cuprates has suggested that an AFM QCP exists in n-doped cuprates. For example, in $(Nd,Ce)_2CuO_4$ (NCCO), ARPES [16,17] and Shubnikov quantum oscillation (QO) experiments [18,19] suggest a FSR at Ce= 0.17. In contrast, a normal state Hall Effect critical doping is reported to be near optimal doping (Ce= 0.145) [20], very close to where the long range order AFM ends, but rather different from where quantum oscillation (QO) experiments suggest that the FS reconstructs (i.e., Ce =0.17). In $(Pr,Ce)_2CuO_4$ (PCCO) the Hall Effect shows a critical doping at Ce= 0.17[21] but no QO or ARPES studies have been done on PCCO over an extended doping range. Also, T-linear resistivity is only found at one doping in NCCO and PCCO. This behavior of NCCO and PCCO is not fully understood and is another significant motivation for our present transport study of LCCO.

Figure 1(a) displays the *ab*-plane resistivity ($\rho_{xx}$) versus temperature T for six LCCO c-axis oriented films at *H*=0. The resistive superconducting transition $T_c$ has the similar trend as reported earlier [3]. Figure 1(b) illustrates the temperature dependent resistivity to show the normal state behavior of *x* = 0.13 and 0.14 compositions at an applied magnetic field of *H* >$Hc_2$. The 13% doped sample shows an upturn at low temperatures starting from 17 K and tends to saturation at low temperatures as observed for other dopings (*x*=0.11, 0.10). The sample 0.08 has an upturn at low temperatures, however it does not saturate at low temperatures unlike the samples 0.10≥x≥0.13 (see supplementary figure S1). The minima of the normal state resistivity at low temperatures are defined as T$\rho_{min}$ (T at ρ minima) shown in figure 1(b). The 14% doped sample does not show any upturn down to 400 mK. As found previously [3], a low temperature T-linear resistivity is found for Ce doping above 0.14 for doping within the superconducting dome. Our similar data for Ce=0.15 and 0.16 is shown in Fig.1(c)

In Fig. 2 we show the normal state Hall coefficient of LCCO films as a function of temperature (measured from 100 K to 1.8 K) for different Ce doping. The absolute value of the Hall coefficient measured at 14 T jumps dramatically between 13 % and 14% doping. The Hall coefficient of the films with doping $x \geq 0.14$ shows a positive value which is constant below 10 K, and there is a sign change at 1.8 K between doping 0.13 and 0.14. The Hall coefficient for samples 0.10 <x<0.13 as a function of temperature shows a peak (T$_{RHmax}$) and starts to fall at a temperature which depends on the doping. The dotted black lines are an extrapolation to T=0 under an assumption of no FSR and that all the samples have behavior similar to the overdoped samples (x≥0.14.)

Figure 3 displays the temperature vs doping (Ce) phase diagram of La$_{2-x}$Ce$_x$CuO$_4$. The hatched regime is the AFM measured by in-plane angular magnetoresistance ending at x=0.14 [8]. The yellow regime is the superconducting dome. The normal state in-plane resistivity minima, T$\rho_{min}$ is determined from the derivative (dρ/dT). The normal state in-plane Hall resistivity maxima, T$_{RHmax,}$ ends at x=0.14. The estimated Fermi surface reconstruction line T$_{FSR}$ (solid blue line) separates the large Fermi surface region from the reconstructed Fermi surface as a function of doping. The dotted blue line is the extrapolation of T$_{FSR}$ assuming that T$\rho_{min}$ is due to only to the FSR.

Figure 4 displays the in-plane electrical resistivity ρ of two LCCO samples as a function of temperature, with doping x as indicated. The red curve is data taken in zero magnetic field (H = 0). The black curve is the fitted data of the red curve using ρ(T)= ρ$_0$+AT$^n$ (ρ$_0$ is the residual resistivity (45 μΩ-cm for 0.11, 23 μΩ-cm for 0.13), n=2 ) above Tc and has been extrapolated to T→0 to get ρ$_0$ assuming there is no upturn. The green line is the normal state resistivity measured at 10 T with ρ(0) (73 μΩ-cm for 0.11, 27 μΩ-cm for 0.13) its extrapolation to T=0.

In electron doped cuprates commensurate (π, π) spin density wave (SDW) order has been detected by muon spin rotation and neutron diffraction [1]. This SDW order (long range or short range) exists over a wide range of doping starting at the undoped AFM state and vanishing at a critical doping $x_c$, where the resistivity minima [21,22] and in-plane angular magnetoresistance also vanish [23]. Theory [24,6] suggests that there should be a quantum critical point separating the overdoped paramagnetic state, with a large Fermi surface, from the SDW state with a reconstructed Fermi surface of small electron and hole pockets. This is experimentally suggested in electron doped NCCO and PCCO near optimal doping by low temperature QO [19,20] and ARPES measurements [17,18]. A Fermi surface reconstruction was also suggested by earlier normal state Hall measurement on PCCO, where an abrupt drop of the Hall coefficient and sign change was found at 300 mK as one approached optimal electron doping from the overdoped side [21].

As shown in figure 2 the normal state Hall coefficient at 1.8 K for LCCO as a function of doping suddenly drops and changes sign between 0.13 and 0.14, which in analogy with PCCO, strongly suggests a Fermi surface reconstruction at x=0.14. The 2D Fermi surface of most cuprates is well established from ARPES and QO experiments. For n-type at higher doping, the FS is a large hole-like cylinder and for underdoped the FS has electron pockets. From theory [26] the Hall number ($n_H=V/eR_H$) in the electron doped cuprates should follow $n_H=1-x$ at doping above SDW reconstruction and $n_H=-x$ for the under doped regime well below the FSR. Our data for LCCO, shown in Fig. 5, is in good qualitative agreement with this, however QO and ARPES experiment have not yet been done on LCCO. This is the same behavior as found recently in hole-doped cuprates at very high magnetic fields, where the Hall coefficient goes from 1+p in the overdoped region to p in the lower doped region [9,11]. This suggested a low temperature (T=0K) FSR at a critical doping of p*, the doping where the pseudogap state ends. Since the FSR in the n-type cuprates is caused by the onset of short range AFM (when coming from the overdoped side), it may well be that a related short range order can reconstruct the Fermi surface in hole-doped cuprates.

As also shown in Fig 5 the Hall number deviates from the 1-x line for the higher-doped samples. The carrier density has been calculated assuming one band transport, which is supported by a linear in field Hall resistivity for over doped and heavily underdoped samples (see supplementary information and Ref 33). But, we can fit the data with $n_H=1-bx$, where b is a correction parameter of 1.74. A possible explanation for the 1-bx behavior is given in the supplementary information (SI). The doping near the FSR (Ce=0.13) gives a very high negative value of $n_H$. But at this doping LCCO has two types of carriers. So we do not expect a simple one carrier $R_H$ for this doping to fit on either line in Fig.5.

The 0.13 sample shows similar temperature dependence for $R_H$ as the overdoped samples above 17.5 K. The Hall coefficient goes through a maximum at 17.5 K (where the short range AFM regime starts for this doping) and starts to drop from positive to negative. The behavior of the Hall coefficient strongly suggests that if there was no Fermi surface reconstruction the Hall coefficient would roughly follow the black dotted line shown in Fig. 2(b). The difference between the black dotted line and the measured solid line is caused by loss of carriers and mobility change below the Fermi surface reconstruction. So one can surmise that the Fermi surface reconstruction starts at temperature 17.5 K for x=0.13. We can use the Hall coefficient maxima as the temperature where the Fermi surface reconstruction starts for each doping (0.11 at 27.5 K and 0.10 at 35 K) as temperature decreases. This low temperature drop of Hall coefficient, seen in samples with x <0.14 can be attributed to the Fermi surface reconstruction due to SDW (AFM) order below $T_{FSR}$ in the hatched regime shown in Fig. 3. All over doped samples (x≥0.14) should have a large hole like-Fermi surface at low temperatures. This needs to be confirmed by ARPES and/or QO experiments in the future.

We now discuss some features of the normal state resistivity. As shown in Fig. 1(c), we find a normal state low temperature linear in T resistivity for a range of Ce doping at, and above, the FSR. Our data here is in accord with resistivity measured previously to even lower temperatures [3]. This is a very anomalous and unexplained resistivity behavior. A T-linear resistivity at the FSR doping can be understood as scattering associated with the fluctuations at temperatures above a QCP, but similar very low temperature behavior at higher doping can not be explained by the usual quantum critical theory [5,6]. Our results suggest that the FSR and the T-linear resistivity are closely connected, but the exact relation is a mystery. A doping range of T-linear resistivity has also been observed in some hole-doped cuprates [25,26] at, and above, the pseudogap end point. However in contrast to n-type LCCO, it has not been possible to apply large enough magnetic fields to probe the normal state at very low temperatures,i.e, to accesss the ground state. Nevertheless, the very similar behavior in electron and hole-doped cuprates, suggests that the close connection between a FSR and T linear resistivity are crucial to understanding the HTSC. The temperature dependent *ab*-plane resistivity exhibits a resistivity minimum at low temperatures for samples x≤0.13 and no minimum for higher doping. This is a well-known feature of all cuprate superconductors. In very under doped (x=0.05 and 0.10) PCCO the low-temperature resistivity upturn was attributed to 2D weak localization [27]. However, the resistivity tends to saturate as the temperature approaches zero for samples near the FSR. This low temperature saturation cannot be explained by 2D weak localization where the resistivity should obey ρ α logT. Later, the upturn observed in PCCO and NCCO was attributed to a Kondo effect due to scattering of conduction electrons by unpaired $Cu^+$ spins [28]. But, Dagan et al.[29] found that for PCCO all doping below the FSR show an anisotropic magnetoresistance. Since this rules out Kondo scattering Dagan et al. suggested another form of spin scattering, with the spin linked to the AFM, as the cause of the upturn. This explanation has received support in a theoretical proposal by Chen et al. [30].

Here, we suggest an alternative explanation for the doping close to the FSR. We note that the $T_{RHmax}$ of the Hall coefficient as a function temperature and Tρmin of the resistivity of LCCO are at the same temperature for the samples 0.010≤x≤0.13 as shown in Fig. 3. This correlation strongly suggests that the low temperature resistivity upturn is due to carrier and mobility changes below the Fermi surface reconstruction. For the doping near the FSR we try an analysis similar to that done recently in hole-doped cuprates [10]. We take 1/ρ=neµ for one carrier transport and we assume that the mobility does not change due to the FSR. As T→0 $n_ρ$(with FSR)/n(without FSR) = $ρ_0$ /ρ(0) where $ρ_0$ is the residual resistance assuming no FSR at T→0 and ρ(0) is the resistivity due to loss of carriers associated with the FSR (see Fig. 4). So $n_ρ$=n($ρ_0$ /ρ(0)). For the large Fermi surface n=1-x, thus $n_ρ$=(1-x)($ρ_0$ /ρ(0)). This $n_ρ$ should be the Hall number below the FSR. The experimental value of $ρ_0$/ρ(0) is 0.62 and 0.85 for x=0.11 and x=0.13 samples respectively. Calculating $n_ρ$ using the above expression gives 0.55 for x=0.11 and 0.74 for x=0.13. The measured values of $n_H$ are 0.13 for x=0.11 and 6.2 for x=0.13, where $n_H$ =V/e$R_H$ (*V* is volume per copper, *e* charge of the carrier and $R_H$ the measured Hall coefficient). If the size of the upturn only depended on the loss of carriers then the values of $n_H$ and $n_ρ$ should be the same (an alternative calculation is shown in the SI). So the resistivity upturn at low temperatures cannot be explained only by loss of carriers. There must also be a mobility change. This experimental result is supported by a recent theory paper from Sachdev's group [31]. This is not at all surprising for the x=0.13 doping since this doping clearly has two types of carriers and cannot be explained by a one band model (see SI). Thus the size of the upturn in the normal state resistivity in electron-doped cuprates is more complex than its counterpart hole-doped materials whose resistivity upturn has been explained only by a drop of carrier density [10].

The low temperature upturn seen in heavily under doped n-type samples cannot be explained by the Fermi surface reconstruction alone. The heavily under doped samples, unlike optimal and slightly under doped samples, do not show a low temperature resistivity saturation as temperatures approaches to zero (see supplementary Fig. S2 and ref 32). The resistivity of these samples is two orders of magnitude higher than optimal or slightly higher doped samples at low temperatures. For these samples the upturn in normal state resisivity is probably a combination of the FSR and disorder localization which gives logarithmic increase of resistivity as temperatures tends to zero.

In conclusion, we have performed low temperature, normal state (H > Hc2), ab-plane resistivity and Hall effect measurements on electron-doped $La_{2-x}Ce_xCuO_4$ as a function of doping. Our results give very strong evidence for a Fermi surface reconstruction (FSR) at x=0.14. The low temperature resistivity shows an upturn below x=0.14 and the Hall number as a function of doping drops at 0.14 from 1-x to -x. The Hall resistivity at $0.18 \geq x \geq 0.14$ and $0.11 \geq x \geq 0.08$ is linear with magnetic field and at x=0.13 becomes nonlinear, more evidence for a change in the Fermi surface and the existence of two types of carriers at this doping. We find a low temperature linear-in-T resistivity for an extended range of doping beyond the FSR doping. This anomalous behavior is unexplained, but it appears to impact the high $–T_c$ superconductivity found in zero magnetic field. The low temperature resistivity upturn found for doping below 0.14 can be explained by a change in carrier number and mobility below the FSR. Our work shows that there are the striking experimental similarities between the transport properties of electron and hole-doped copper oxides and provides evidence that the normal state near the FSR doping is similar in all the cuprates. The cause of the FSR is a commensurate spin density wave in the n-doped cuprates but is yet to be determined in the hole-doped cuprates.

Acknowledgement: We thank L. Taillefer for valuable discussions. This research was supported by the NSF under DMR-1410665, the Maryland "Center for Nanophysics and Advanced Materials" and National Institute of Standards and Technology, U.S. Department of Commerce, in supporting this research through awards 70NANB12H238 and 70NANB15H261.

.

**Figure Captions**

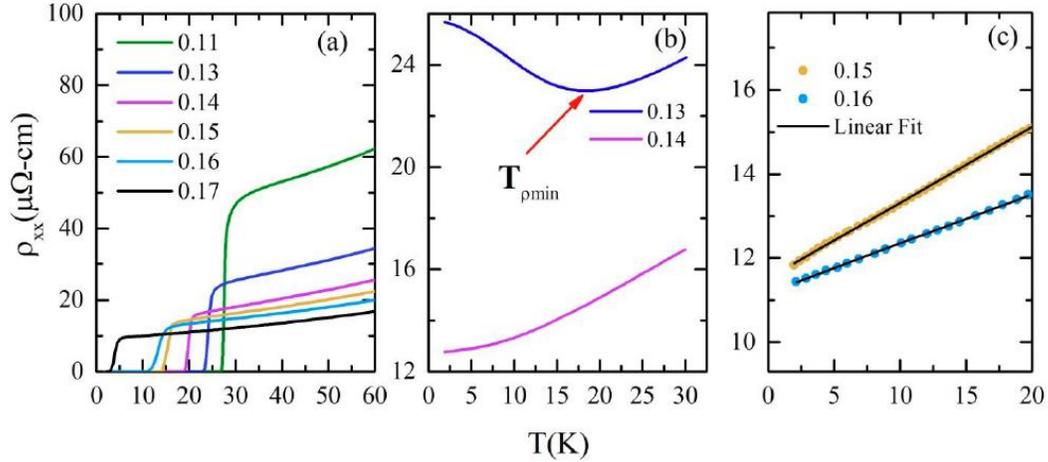

**Figure 1.** (a) (color online). *ab*-plane resistivity versus temperature for $La_{2-x}Ce_xCuO_4$ films with various Ce doping. (b) The normal state *ab*-plane resistivity versus temperature in a magnetic field of $H > Hc_2$ applied parallel to the *c* axis for x=0.13 (8 T) and x=0.14 (6 T). ). (c) Normal state resistivity below 20 K for x=0.15 and 0.16 with linear fit.

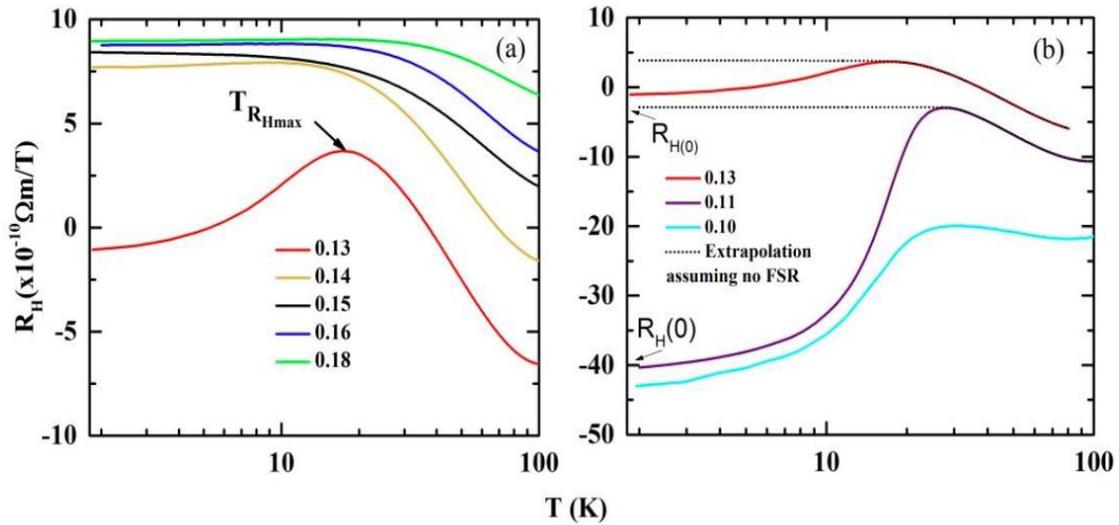

**Figure 2**. (a), (b). The Hall coefficient versus temperature for $La_{2-x}Ce_xCuO_4$ films with various Ce doping (*x*) measured at a magnetic field of 14 T (solid lines). The dotted black lines are an extrapolation assuming no Fermi surface reconstruction (FSR).

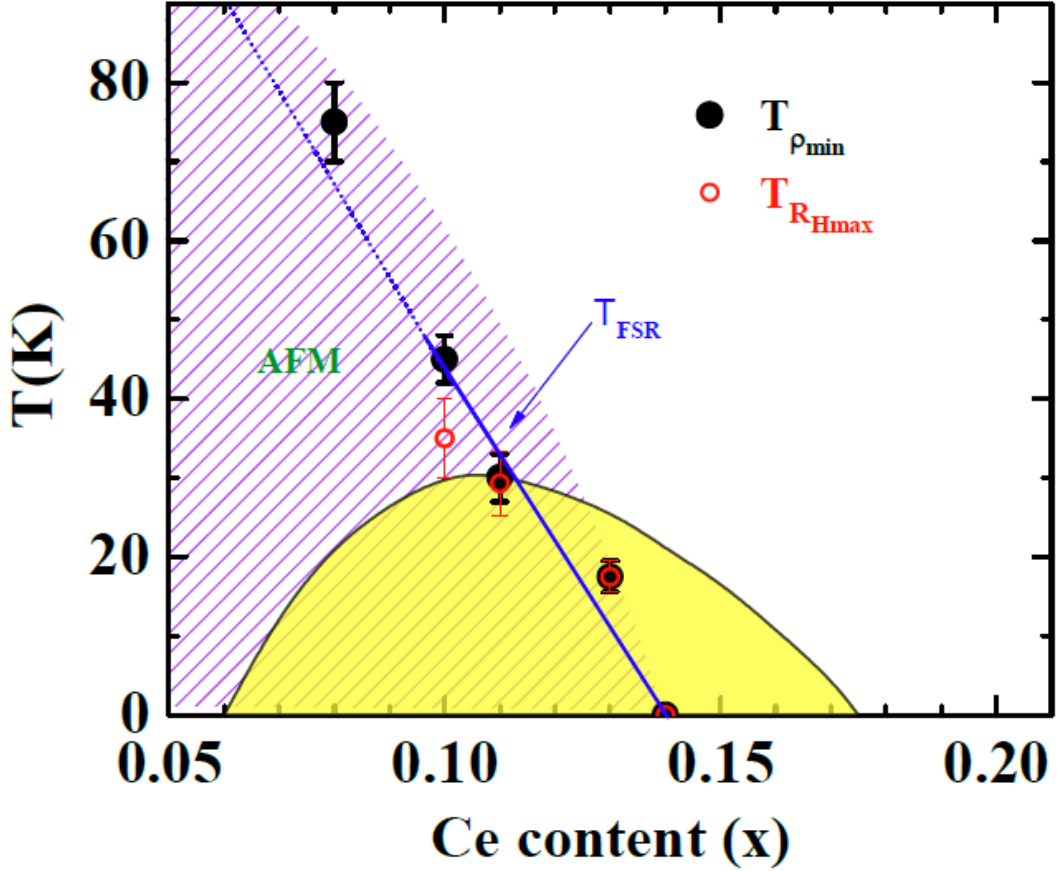

**Figure 3.** Temperature vs doping (Ce) phase diagram of $La_{2-x}Ce_xCuO_4$. The hatched regime is the AFM region measured by in-plane magnetoresistance ending at x=0.14 (ref-[8]). Yellow regime is the superconducting dome. $T\rho_{min}$ (Black filled circle)) is the normal state in-plane resistivity minima ending at x=0.14. $T_{RHmax}$ (Hollow red circle) is the normal state in-plane Hall resistivity maxima ending at x=0.14. $T_{FSR}$ is the Fermi surface reconstruction line (solid blue line) which separate the large Fermi surface from the reconstructed Fermi surface. Dotted blue line is the extrapolation of $T_{FSR}$.

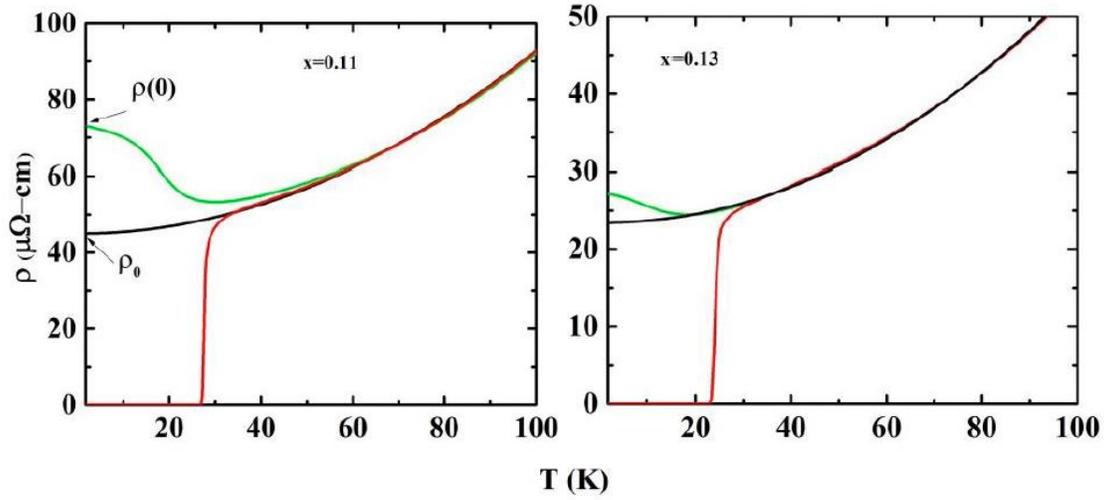

**Figure 4.** In-plane electrical resistivity (ρ) of two LCCO samples as a function of temperature, with doping *x* as indicated. The red curve is data taken in zero magnetic field ($H = 0$). The black curve is the fitted data of The red curve above Tc and is extrapolated to T→0 to get $\rho_0$. The green line is the normal state resistivity measured at 10 T with ρ(0) the normal state resistivity at T→0.

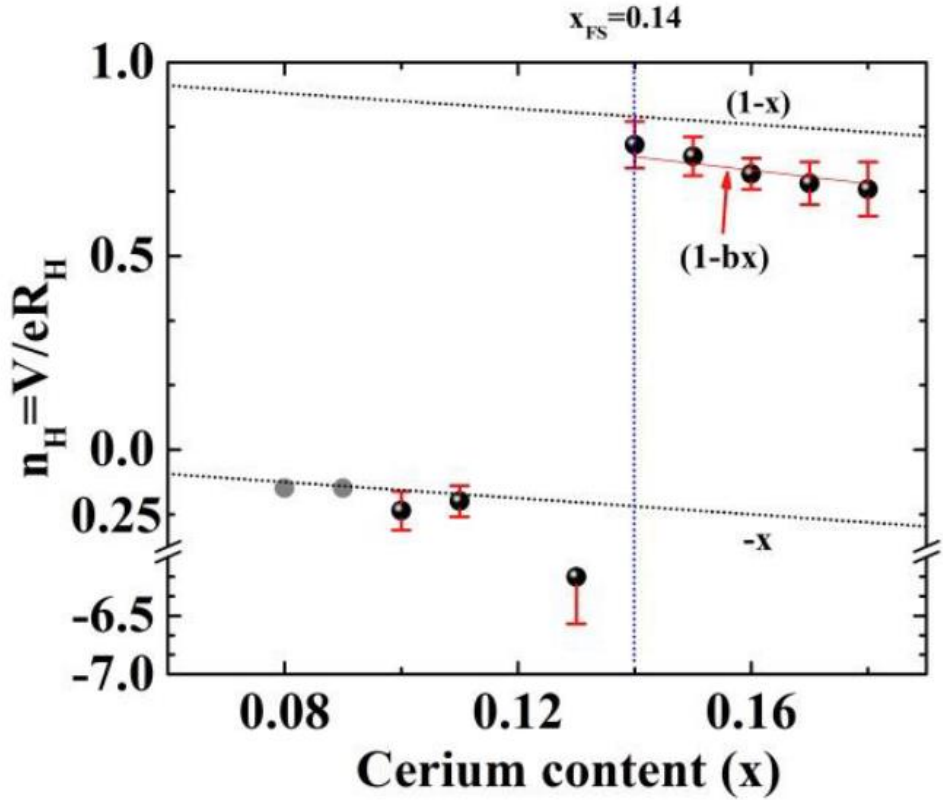

**Figure 5.** Hall number $n_H=V/eR_H$ at 1.8 K as a function of Ce doping with single carrier fitting $n_H=1-x$ and $n_H=-x$. Red solid line is the $n_H=1-bx$ fitting where b is a fitting parameter. The gray data points of 0.08 and 0.09 Ce doping are taken from (ref [33]). Error bars are coming from the error in the film thickness measurement.

**Supplementary Information**

**Experimental methods**

We have measured seven La$_{2-x}$Ce$_x$CuO$_4$ films to cover under-doped (x=0.08, 0.10), optimally doped (x=0.11) and overdoped (x=0.13, 0.14, 0.15, 0.16, 0.17, and 0.18) compositions. High quality La$_{2-x}$Ce$_x$CuO$_4$ films with thickness about 150 - 200 nm were grown by pulsed laser deposition (PLD) on SrTiO$_3$ [100] substrates (5×5 mm$^2$) at a temperature of 700 °C and at an oxygen partial pressure of 230 mTorr. The quality of the samples was determined based on: a) The superconducting transition width ($\Delta$Tc) calculated from the imaginary part of the AC susceptibility peak, b) lowest residual resistivity of the samples and c) highest Tc of the samples. The transition width (full width at half maximum of the peak in d$\rho_{xx}$/dT) of the films is within the range of 0.2-0.8 in the optimum and overdoped films. The width increases with increasing Ce doping concentration. The PLD targets were prepared by the solid-state reaction method using 99.999% pure La$_2$O$_5$, CeO$_5$, and CuO powders. The thin films were characterized by X-ray diffraction and ac susceptibility measurements. The crystalline phase was determined by Bruker X-ray diffraction (XRD) which showed *c*-axis oriented epitaxial LCCO tetragonal phase. The thickness of the films was determined by cross sectional scanning electron microscopy (SEM). The Hall coefficient was measured by applying a magnetic field (14 T) perpendicular to the film plane in the Hall bar geometry (with In-Ag soldering) in a Physical Property Measurement System (PPMS) equipped with a 14 T superconducting magnet. The magnetic field was swept from 14 to − 14 T under a constant current. The MR contribution was subtracted in the R$_H$ measurement using positive and negative magnetic field. The Hall bar geometry was made by photolithography followed by the Argon ion milling. The transport properties were measure from 100 K to 1.8 K. Some examples are shown here in Fig S1 and Fig-1 in the main text.

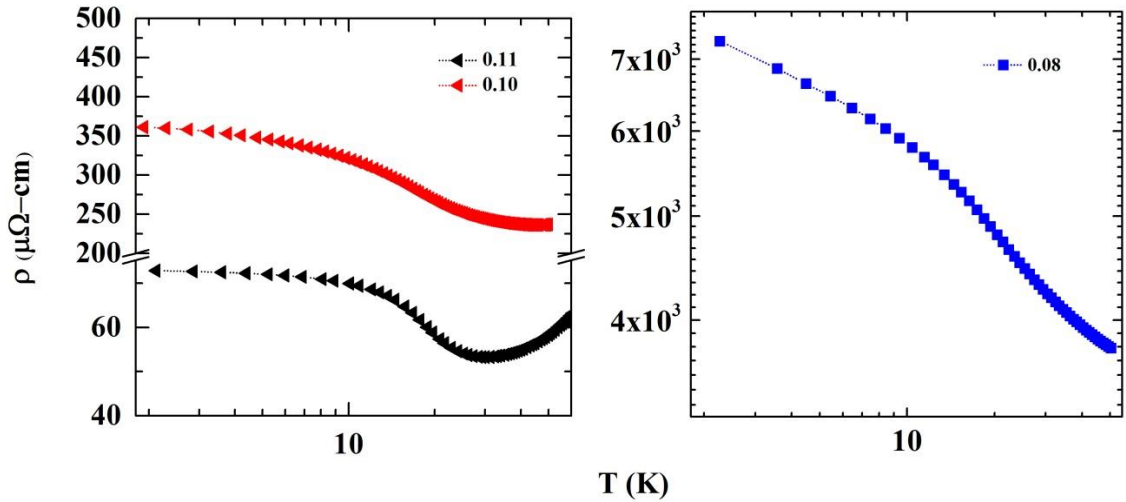

Fig S1. The normal state *ab-plane* resistivity of LCCO x=0.11, 0.10 and 0.08 as a function of temperature. Normal state resistivity of the samples 0.11 and 0.10 are measured at the magnetic field 10 T and for 0.08 at 6 T.

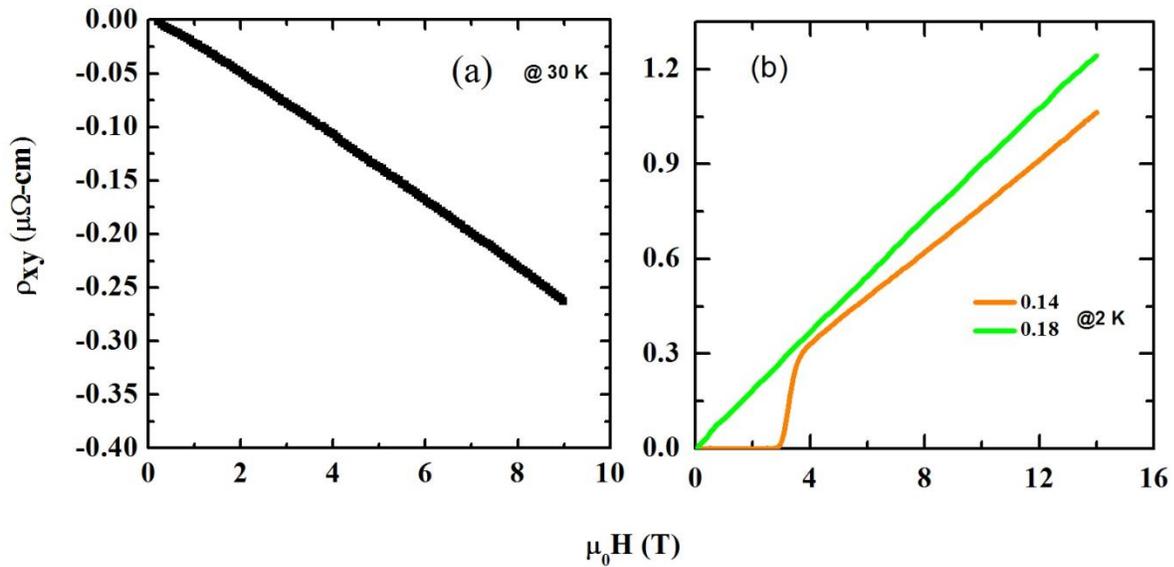

Figure S2. The Hall resistivity as function magnetic field (a) the Hall resistivity of x= 0.11 measured at 30 K. Low temperatures data is not shown here due to very small range of normal state which makes it very difficult to judge the linear behavior. Figure S2 (b) shows the Hall resistivity as a function of magnetic field measure at 2 K.

A linear Hall resistivity as a function of field is found in the samples 0.18≥x≥0.14 (see Fig S2). This behavior can be can be understood based on a one band Drude model with the charge transport solely due to holes in overdoped samples. This implies there is only one large hole like Fermi surface above x≥0.14. Under doped samples show a negative linear Hall resistivity which suggests that they are dominated by electron pockets of the reconstructed Fermi surface.

**Non-linear Hall resistivity**

To understand the nonlinear Hall resistivity seen in the x=0.13% sample one has to consider two band conduction. Here we have taken classical two band Drude model [1] to understand the behavior of the Hall resistivity in x=0.13 samples considering the Hall resistivity arises due to competing contribution from electron and hole like orbits in the reconstructed Fermi surface. Assuming two carrier conduction, the Hall resistivity of the metal can be modeled as [1,2]

$$\rho_{xy}(H) = \frac{(\sigma_e^2 R_e + \sigma_h^2 R_h + \sigma_e^2 \sigma_h^2 R_e R_h (R_e + R_h) H^2)}{(\sigma_e + \sigma_h)^2 + \sigma_e^2 \sigma_h^2 (R_e + R_h)^2 H^2} H \ldots (1)$$

$$\rho_{xx}(H) = \frac{(\sigma_e + \sigma_h) + \sigma_e \sigma_h (\sigma_e R_e^2 + \sigma_h R_h^2) H^2}{(\sigma_e + \sigma_h)^2 + \sigma_e^2 \sigma_h^2 (R_e + R_h)^2 H^2} \ldots \ldots \ldots (2)$$

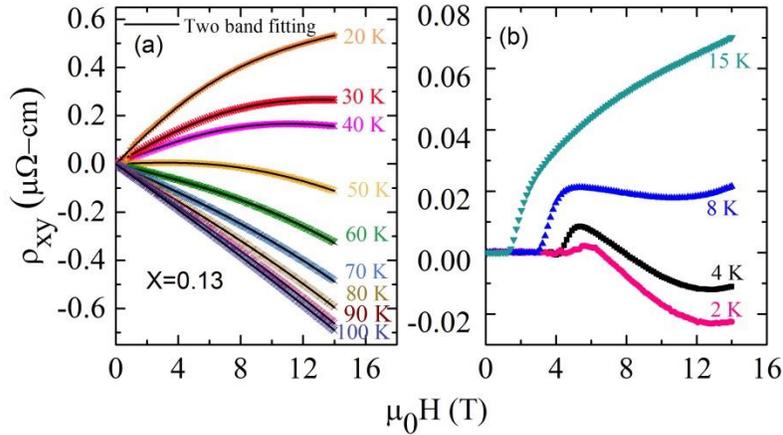

Figure S3. (a) Color scatter lines are hall resistivity at temperatures from 100 K to 20 K with magnetic field for La$_{2-x}$Ce$_x$CuO$_4$ (x=0.13), solid black line is the two carrier Drude model fitting (b) Hall resistivity for x=0.13 temperatures from 15 K to 2K.

where $\sigma_{e(h)}$ and $R_{e(h)}$ are the conductivity and Hall coefficient of the electron(e) and hole(h) carriers, respectively. Based on this equation, we have fitted the $\rho_{xy}(H)$ data with the constraint of

$$\rho_{xx}(0) = \frac{1}{\sigma_e + \sigma_h} \ldots \ldots \ldots \ldots \ldots \ldots \ldots \ldots \ldots \ldots \ldots \ldots \ldots . (3)$$

where $\rho_{xx}(0)$ is the zero field normal state resistivity. The quality of the fitting from temperatures 20 K to 100 K is remarkably good as shown in figure 3(a). The fitting parameters derived from the fittings using Eqs. (1) are shown in figure S4. This two band fitting confirms there are two types of carriers (hole and electrons) in the 13% Ce doped sample. This confirms that a small hole- like Fermi pocket exists in the normal state at low temperatures below 17.5 K. Now as we increase the doping, the small electron pockets in 0.13% vanish at 0.14% at 1.8 K.

Figure S4. shows the fitting parameters derived from the classical two band Drude model in Ce doped 13% LCCO. In the Fig. S4 the value of $R_e$ around 60 K is 18 mm$^3$/C and increases with temperature. The absolute value of zero for the Hall coefficient is unphysical. It can be explained by absence of electron pocket when Re=0 and the absence of the hole pocket when $R_h$=0. From the figure S3 one can say above 60 K towards 100 K the Hall pocket is vanishing. The Hall resistivity vs magnetic field data shows almost linear behavior at higher temperatures (100 K). So the Hall conductivity becomes zero. However below 60 K to 20 K it is dominated by holes. But there is still a small electron pockets coexist as one can see the electron conductivity do not go zero. Below 20 K, we believe the hole pocket gets smaller and electron becomes dominant as the figure 2 in the main text shows the Hall resistivity in the normal state changes sign from positive to negative. However, the fitting for temperatures below 20 K was not possible due to complex Hall resistivity characteristic with the magnetic field due to mixed state vortex motion as well as the spin scattering in the normal state resistivity. The spin scattering below 20 K (where the upturn starts in the normal state resistivity) makes it difficult to get normal state resistivity value. The analogy is drawn here based on recently published paper on YBCO by Rourke et. al [1]. However the temperature dependent Hall coefficient might be more complex than what we have interpreted here. One important message we can conclude that the 13% doped Ce doped LCCO has small hole carrier contribution with the electron carrier at the lowest temperature measured at 1.8 K

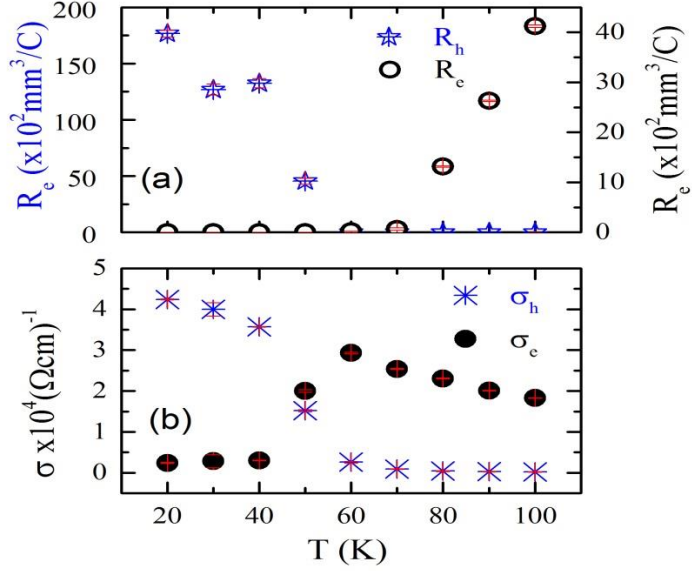

Figure S4. (a) $R_h$ and $R_e$ Hall coefficients of hole and electron (b) $\sigma_h$ and $\sigma_e$ electrical conductivity of hole and electron derived from the fits of two band Drude's model (x=0.13).

**An alternate estimate for the resistivity upturn below the FSR**

Here we show an alternate calculation to correlate the change of the resistivity with the drop of Hall coefficient (difference between the dotted black line at T→0 ($R_{H(0)}$) assuming no FSR and measured solid line T→0 ($R_H(0)$)as shown in Fig. 2). If we consider the change in the resistivity is only due to a loss of carriers then, $1/\rho=ne\mu=V/\mu R_H$. So $R_{H(0)}/R_H(0)= \rho_0 /\rho(0)$. The value of $R_{H(0)}/ R_H(0)$(0.067 for x=0.11) is one order smaller than the $\rho_0 /\rho(0)$ (0.616). This also shows that the resistivity upturn at low temperatures cannot be explained only by loss of carriers.

**The plausible explanation for the deviation of $n_H$=1-x to 1-bx**

As also shown in Fig 5 (main text) the Hall number deviates from the 1-x line for the higher-doped samples. The carrier density has been calculated assuming one band transport, which is supported by a linear in field Hall resistivity for over doped and heavily underdoped samples (see S2 and Ref 3). But, we can fit the data with $n_H$=1-bx, where b is a correction parameter of 1.74. We have defined the b as a correction factor in the doping concentration. In electron-doped cuprates the doping dependence depends on the Ce content and the oxygen content. To achieve the optimal properties the n-type cuprates are annealed in vacuum, which can create oxygen vacancies, so that $La_{2-x}Ce_xCuO_4$ should really be written as $La_{2-x}Ce_xCuO_{4-\delta}$. Hence, we are changing two parameters to get the optimal superconductivity. Any change in $\delta$ will affect the true carrier concentration. The oxygen vacancy effectively adds electrons to the system, i.e, x become bx. So the actual doping in the system could be higher than that of the Ce content (x). We take b as a correction factor to the carrier density due to any contribution from oxygen vacancies.

The difference between $n_H$ (=1-x) and $n'_H$(=1-bx) is the change in the Hall number due to oxygen vacancies. Now if we take the 15 % sample to calculate the difference in Hall Number we find $\Delta n_H=(n_H- n'_H)=0.11$. If we convert this number to a change in Hall coefficient we find $\Delta R_H=(R'_H-R_H) =(V/en'_H - V/en_H)= 1.0 \times 10^{-10} \Omega m/T$. Is this reasonable? Higgins et al. [4] reported that changes in the oxygen content in over doped Ce=0.17 PCCO can change the value of $R_H$ from $5.5 \times 10^{-10}$ to $7.5 \times 10^{-10}$ ($\Omega m/T$), which is about 2 times higher than what we estimate for LCCO. Thus it is quite reasonable that our change in carrier number 1-x to 1-bx could be caused by oxygen vacancies.

Another possible origin of the deviation from 1-x carrier number is the shape of the Fermi surface for doping above the FSR. In the theory of Lin and Millis [5] for the Hall effect of n-type cuprates they found the Fermi surface shape could affect the value of the Hall number, but not the slope b. Our data suggests that the oxygen deficiency is the more likely explanation for the deviation in $R_H$ at higher doping. The Hall number for under doped samples $0.08 \leq x \leq 0.11$ follows $n_H=-x$. The deviation of the measured Hall coefficient from the $n_H =-x$ line is negligibly small, i.e, no oxygen vacancy correction needed. The reason for this is not clear, but it could be that below the FSR the oxygen vacancy formation energy is higher when electron carriers are dominant. The doping near the FSR (Ce=0.13) gives a very high negative value of $n_H$. But at this doping LCCO has two types of carriers. So we do not expect a simple one carrier $R_H$ for this doping to fit on either line in main text Fig.5.